\documentclass[11pt,twoside]{article}

\usepackage{asp2006}
\usepackage{epsf}
\usepackage{lscape}
\usepackage {graphics}
\usepackage{graphicx}

\DeclareGraphicsExtensions{.jpg, .eps,.ps}

\newcommand{\Msun}{M_\odot}

\begin{document}
\title{Simulating the Gaseous Halos of Galaxies}
\author{Tobias Kaufmann\altaffilmark{1}, James S. Bullock\altaffilmark{1}, Ari Maller\altaffilmark{2} and Taotao Fang\altaffilmark{1} } 
\altaffiltext{1}{Center for Cosmology, Department of Physics and Astronomy, University of California, Irvine, CA, 92697} 
\altaffiltext{2}{Dept. of Physics, New York City College of Technology, CUNY, NY, 11201}   

\begin{abstract}

Observations  of local X-ray absorbers, high-velocity clouds,
and distant quasar absorption line systems  suggest    that a significant fraction  of
 baryons may reside in multi-phase, low-density, extended, $\sim 100$  kpc,  gaseous halos around normal galaxies.   We present a pair of high-resolution SPH (smoothed particle hydrodynamics)
 simulations that explore the nature of cool gas infall into galaxies, and the physical
 conditions necessary to support the type of gaseous halos that seem to be required by observations.  The two simulations are identical other than their initial gas density distributions:
one is initialized with a {\em standard} hot gas halo that traces the cuspy profile of the dark matter, and the other
is initialized with a {\em cored} hot halo with a high central entropy, as might be expected in models with early pre-heating feedback.  Galaxy formation proceeds in dramatically
different fashions in these two cases.  While the standard cuspy halo cools rapidly, primarily from the central region, 
the  cored halo is quasi-stable for $\sim 4$ Gyr and eventually
cools via the fragmentation and infall of clouds from $\sim 100$ kpc
distances.   After 10 Gyr of cooling, the standard halo's X-ray luminosity  is $\sim 100$ times current limits and the resultant disk galaxy  is twice as massive as the Milky Way.  In contrast, the cored halo
has an X-ray luminosity that is in line with observations, an extended cloud population reminiscent of the high-velocity cloud population of the Milky Way, and  a disk galaxy with half the mass and $\sim 50\%$ more specific angular momentum than the disk formed in the low-entropy simulation.   
These results suggest that  the distribution and
character of halo gas provides an important testing ground for galaxy formation models
and may be used to constrain
the physics of galaxy formation.

\end{abstract}


\section{Introduction}

It is well known that in the absence of feedback
the  majority  of  baryons in galaxy-size    dark matter halos  ($M_{\rm v} \sim
10^{12}$  M$_\odot$)  should have  cooled into halo  centers
over a     Hubble  time (e.g. White \& Rees 1978;  Katz et al. 1992; Benson et al. 2003).
In
contrast, only $\sim  20 \%$ of  the  associated baryons  in Milky-Way
size halos are observed to be in a cold, collapsed form (Maller \& Bullock 2004; Mo et al. 2005; Fukugita \& Peebles 2006; Nicastro et al. 2007).
An understanding of the 
feedback processes that  act  to  solve this   galaxy  ``overcooling''
problem is a major goal of galaxy formation today.  
It  is not known  if the  unaccounted baryons exist  primarily as
plasmas within normal  galaxy halos  (Maller \& Bullock 2004; Fukugita \& Peebles 2006; Sommer-Larsen 2006) or if they  have   been largely expelled as a result of energetic blow-out (see, e.g., Oppenheimer \& Dav{\'e}   2006, for a discussion).

Observations  of local X-ray absorbers (Williams et al. 2005; Fang et al. 2006), 
high-velocity clouds (Collins et al. 2005; Thom et al. 2006; Peek et al. 2007), 
and distant quasar absorption line systems (Tumlinson \& Fang 2005; Kacprzak et al. 2007; Tinker \& Chen 2007), suggest    that a significant fraction  of
the missing halo baryons may reside in multi-phase, extended, $\sim 100$  kpc,  gaseous halos of normal galaxies.   However, 
any hot gas around disk galaxies must be
 relatively low density in order to evade X-ray emission bounds 
 ($S_x < 10^{-14}$ erg cm$^{-2}$ s$^{-1}$ arcmin$^{-2}$; Rasmussen et al. 2006, Li et al. 2007).
These results, together with the fairly high covering factors in cool clouds
implied by absorption line studies  ($\sim 50 \%$; Kacprzak et al. 2007), 
are suggestive of a model where normal galaxies are surrounded by extended, low-density 
hot ($\sim 10^6$ K) halos that are filled with fragmented, pressure supported cool ($\sim 10^4$ K) clouds (Maller \& Bullock 2004).

Independently, models aimed  at explaining the optical properties of
galaxies  have relied    increasingly  on the   idea  that  extended,
quasi-stable hot gas halos develop around massive galaxies
(Kere{\v s} et~al. 2005; Bower et al. 2006, Croton et al. 2006).
It is suggested that these hot halos may be quite susceptible to feedback
mechanisms, which could stabilize the systems to cooling  and help
explain  the observed  bimodality in galaxy
properties (Dekel \& Birnboim 2006).   Observational probes of the gaseous halos of galaxies
provide a potential means of testing these ideas.  Entropy injected from feedback mechanisms
will alter the density distribution of halo gas and affect associated
cooling rates (and thus X-ray emission) and the distribution of cooling clouds
fragmenting within the hot halos. Similarly,
early feedback or {\em pre-heating} before the halo
collapses can affect halo gas profiles in a related
manner, with positive consequences for galaxy properties at $z=0$ 
(Mo \& Mao 2002; Oh \& Benson; Lu \& Mo 2007).

\begin{figure*}
\center
\includegraphics[height=50mm]{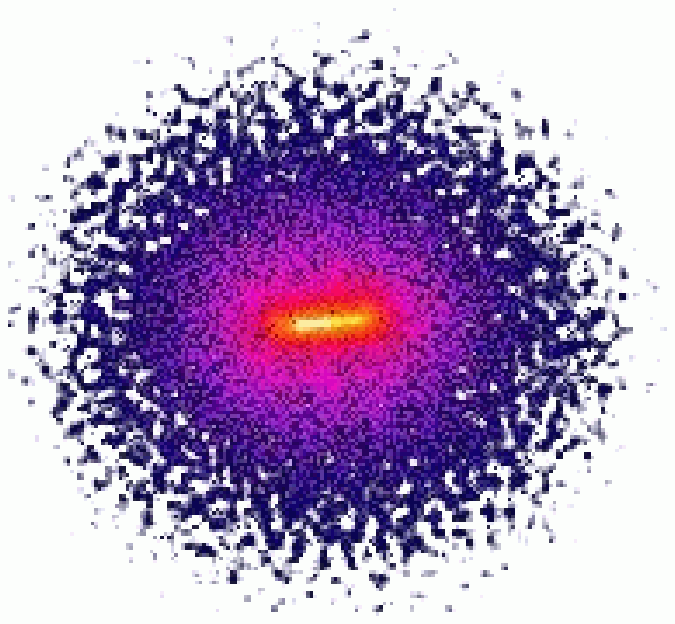}
\includegraphics[height=50mm]{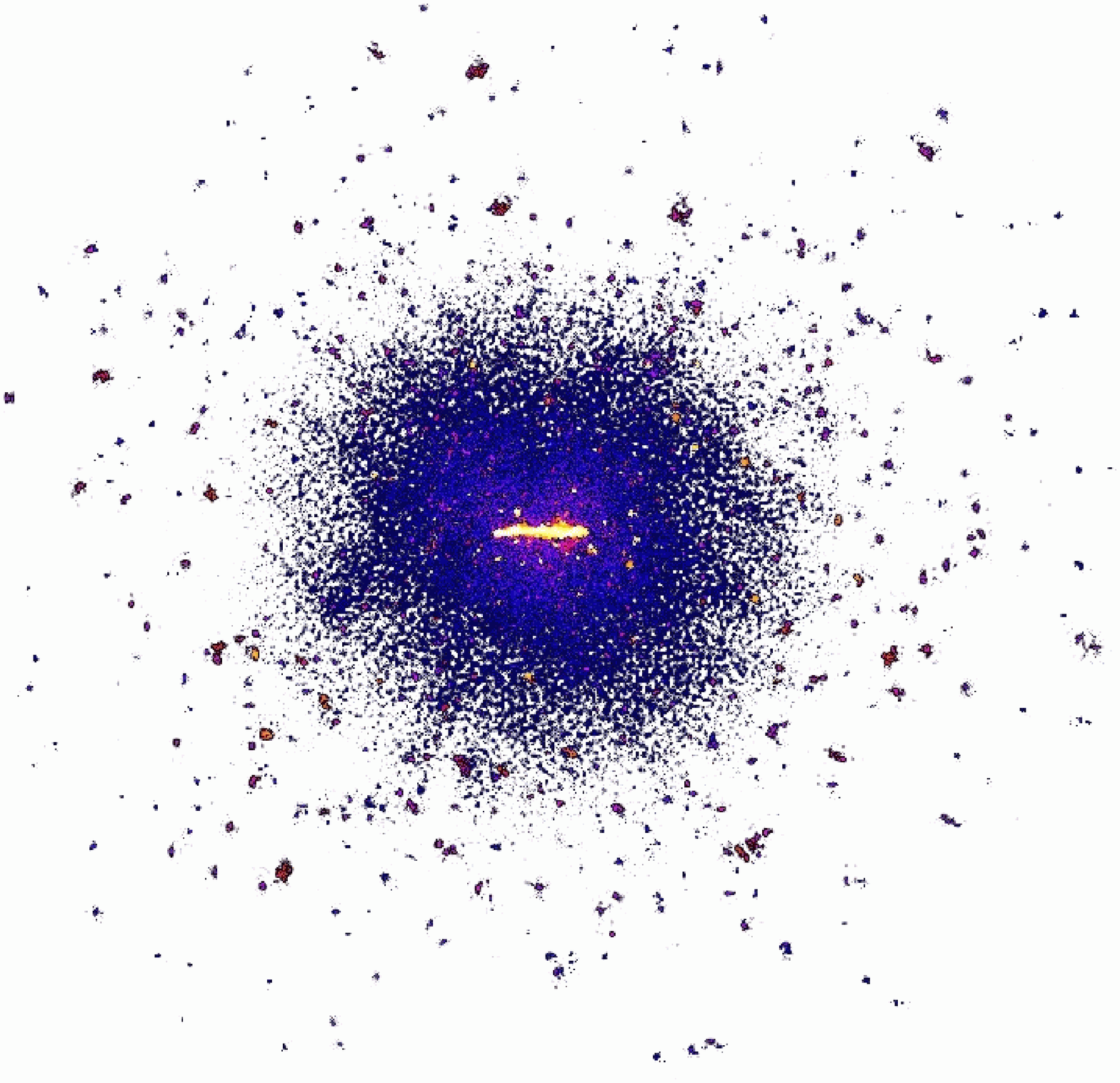}
\caption{Renderings of the total gas density after 10 Gyr of cooling in our simulations. 
{\bf Left:} results for our "standard" halo, initialized with a low-entropy cuspy hot halo.  
{\bf Right:} results for the "cored" halo, initialized with a high-entropy, low-density hot halo
of the same mass. 
In the first case, gas is concentrated around the galaxy.  In the second case, the 
dense gas is more extended and distinct clouds are visible out to $\sim100$ kpc.
The boxes are $400$ kpc on a side. \label{pic} } 
\end{figure*}

\section{Simulations}

Here we present two high-resolution hydrodynamic simulations aimed studying the
stability of hot gaseous galactic halos and the nature of gas cooling into galaxies.  
The first {\em standard} case is initialized with a  cuspy hot halo profile that traces that of its dark matter halo.  The second {\em cored} case is initialized with large,
low-density core.  The cored model has a high central entropy,  $S_0 = T_0/n_0^{2/3} \simeq  20$ keV
cm$^2$,  of the type suggested  in   scenarios with   substantial
pre-heating (e.g. Mo \& Mao 2002).  
The two simulations are identical other than their initial gas density distributions.
Following Kaufmann et al. (2007), we initialize each system  to be
in hydrostatic
equilibrium with an adiabatic equation of state within NFW (Navarro et al. 1996) dark matter halos of mass $M_v = 10^{12} M_\odot$.   The total gas mass inside the virial radius is each case is $10^{11} M_\odot$.
 We use up to $N=2 \times 10^6$ gas and dark matter particles and impose initial gas angular momentum with  a spin parameter $\lambda =  0.03$.

We track 
 cooling in these halos using the parallel TreeSPH code \textsc{Gasoline} (Wadsley et
al. 2004), which is an extension of the pure N-Body gravity code
\textsc{Pkdgrav} developed by Stadel (2001). The adopted star formation recipe  is as described   in Katz (1992), although we use a higher star formation efficiency factor to limit the computational expenses. We do not include any feedback associated with star formation in these exploratory simulations. 

\begin{figure*}[t!]
\center
\includegraphics[scale=0.7]{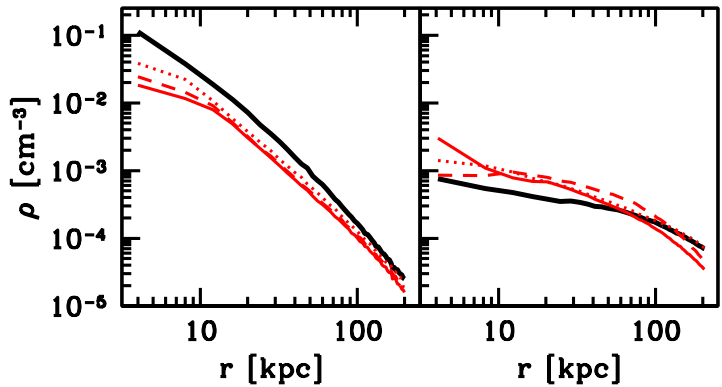}
\caption{Evolution of the hot gas density, standard (left) and cored model (right). The hot halo of the cored model remained relatively stable for  almost $10$  Gyr. Line style: initial conditions (thick solid), evolution with cooling after 3 (dotted), 7 (dashed) and 10 (solid) Gyr. \label{dens_comp} } 
\end{figure*}

\section{Results and qualitative comparisons with observations}

As illustrated in Figures 1, 2, and 3, the different initial density distributions lead to a completely different cooling behaviors. The cuspy halo model cools quickly from the central region
and forms a massive, $6 \times 10^{10}$ $\Msun$, disk galaxy after 
10 Gyr.  Over the same period, the gas density of the cored model remains
remarkably stable (Figure 2).   Galaxy formation in this case
 proceeds after $\sim 4$ Gyr of quasi-stability
via the infall of cool, fragmented clouds as a result of the thermal instabilities
(Kaufmann et al. 2006; Maller \& Bullock 2004).
The final disk in the cored case is significantly less massive $\sim 3 \times 10^{10}$ $\Msun$
and also has $\sim 50 \%$ more specific angular momentum than in the standard run.

As can be seen in Figures 1 and 4, the residual baryonic halos are also substantially
different in these models.  The "cored" halo produces an extended $\sim 100$ kpc
distribution of fragmented, cool/warm $T \sim 10^4$ K clouds with a total
mass of $\sim 5 \times 10^{9} \, \Msun$, while the
"standard" case yields virtually no extended cool halo.\footnote{Note that we find that the
final {\em mass} of this cool halo component 
is independent of our numerical resolution, as is the overall time evolution history.}  
As shown in left two panels of Figure \ref{mapH}, the cored run yields an extended halo of
cool fragmented gas, with a fairly high covering factor,  as suggested by quasar absorption system studies (e.g.  Kacprzak et al. 2007).
Moreover, the X-ray emission of the cored run lies within the limits mentioned above, while the standard run is $\sim 100$ times too bright in X-ray.
This difference is illustrated by the two right panels of Figure \ref{mapH}, where
we show the X-ray surface brightness calculated using  the MEKAL software
package.~\footnote{see http://heasarc.gsfc.nasa.gov/docs/xanadu/xspec/ } 
A more complete description  will be given in Kaufmann et al. (in preparation).

\begin{figure*}
\center
\includegraphics[scale=0.7]{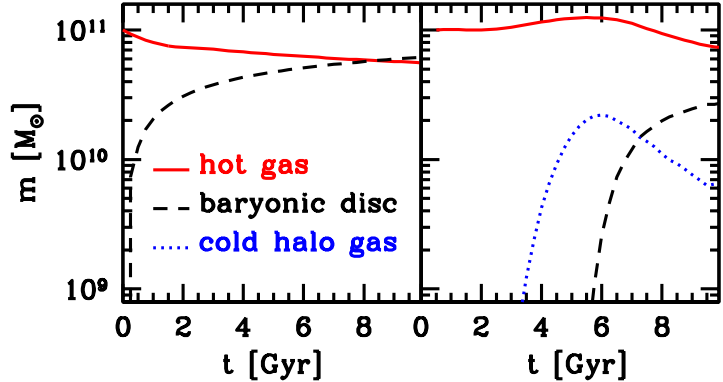}
\caption{Mass evolution of the different gas phases, standard (left) and cored model (right). The cored model ends up with a lighter disk, warm clouds and a more massive hot gas component. \label{massevo_comp} } 
\end{figure*}

\begin{figure*}
\center
\includegraphics[scale=0.45]{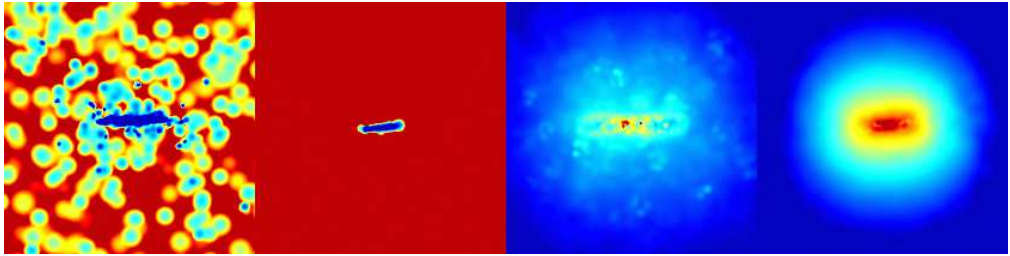}
\caption{
{\bf Left:} Projected density of cool/warm $T< 10^{4.4}$ K gas for the {\em cored} model (left) and {\em standard} cuspy model (right).  The color map spans
$10^{18-21}$ atoms cm$^{-2}$.
{\bf Right:} X-ray surface brightness map 
for the {\em cored} (left) and {\em standard} cuspy model (right).
The color map spans ($10^{-17.1} - 10^{-12.1}$) erg s$^{-1}$ cm$^{-2}$ arcmin$^{-2}$.\label{mapH} } 
\end{figure*}

\section{Summary and Conclusions}

Testing the effects of different initial conditions can teach us about the types of gaseous halos that are required to match the observed distributions of halo gas around galaxies.   These preliminary simulations  show that a cored initial gas density profile with a high initial entropy (as expected in pre-heating scenarios) but without any other feedback 
can produce disk masses, cool gas distributions,  and X-ray emission signals that are
in better agreement with observations than a more standard cuspy halo case with low
central entropy.  These results suggest that  the distribution and
character of gaseous galactic halos provide a powerful tool for understanding
the physics of galaxy formation.

\acknowledgements 
It is a pleasure to thank James Wadsley, Joachim Stadel and Tom Quinn for making
\textsc{Gasoline} available to us. The numerical simulations were performed on the
 IA64  Linux  cluster at the  San Diego Supercomputer Center.    This work was
supported by the Center for Cosmology at UC Irvine.


\end{document}